\documentclass{article}
\usepackage{amssymb}
\usepackage{amsfonts}
\usepackage{amsmath}
\numberwithin{equation}{section}
\DeclareMathOperator{\sgn}{\rm sgn}
\DeclareMathOperator{\diag}{\rm diag}

\DeclareMathOperator{\Rs}{\mathbb{R}}
\DeclareMathOperator{\Cs}{\mathbb{C}}
\DeclareMathOperator{\Go}{\mathcal{G}}

\DeclareMathOperator{\No}{\mathcal{N}}
\DeclareMathOperator{\Do}{\mathcal{D}}
\DeclareMathOperator{\Vo}{\mathcal{V}}

\DeclareMathOperator{\bk}{\mathbf{k}}

\begin{document}

\title{Green's function of heat operator with pure soliton potential}
\author{M.~Boiti${}^{*}$, F.~Pempinelli${}^{*}$, and A.~K.~Pogrebkov${}^{\dag}$ \\
${}^{*}$Dipartimento di Fisica, Universit\`a del Salento and\\
Sezione INFN, Lecce, Italy\\
${}^{\dag}$Steklov Mathematical Institute, Moscow, Russia}
\date{MSC: 37K10, 37K15, 35C08, 37K40\\
Keyword: Kadomtsev--Petviashvili Equation, Heat Operator, Green's Function, Solitons}
\maketitle

\begin{abstract}
The heat operator with a pure soliton potential is considered and its
Green's function, depending on a complex spectral parameter $\bk$, is
derived. Its boundedness properties in all variables and its singularities
in the spectral parameter $\bk$ are studied. A generalization of the Green's function, the extended resolvent, is also given.
\end{abstract}

\section{Introduction}

The Kadomtsev--Petviashvili equation in its version called KPII
\begin{equation}
(u_{t}-6uu_{x_{1}}+u_{x_{1}x_{1}x_{1}})_{x_{1}}=-3u_{x_{2}x_{2}},\label{KPII}
\end{equation}
where $u=u(x,t)$, $x=(x_{1},x_{2})$ and subscripts $x_{1}$, $x_{2}$ and $t$ denote partial derivatives, is a (2+1)-dimensional generalization of the celebrated Korteweg--de~Vries (KdV) equation. The KPII equations, originally
derived as a model for small-amplitude, long-wavelength, weakly two-dimensional waves in a weakly dispersive medium~\cite{KP1970}, have been known to be integrable since the beginning of the 1970s~\cite{D1974,ZS1974},
and can be considered as a prototypical (2+1)-dimensional integrable equation.

The KPII equation is integrable via its association with the operator
\begin{equation}
\mathcal{L}(x,\partial_{x})=-\partial_{x_{2}}+\partial_{x_{1}}^{2}-u(x),\label{heatop}
\end{equation}
which define the well known equation of heat conduction, or heat equation for short. The spectral theory of the operator (\ref{heatop}) was developed in \cite{BarYacoov}--\cite{Grinevich0} in the case of a real potential  $u(x)$ rapidly decaying at spatial infinity, which, however, is not the most interesting case, since the KPII equation was just proposed in \cite{KP1970} in order to deal with two dimensional weak transverse perturbation of the one soliton solution of the KdV.

A spectral theory of the KPII equation that also includes solitons has to be build using the resolvent approach, as already successfully done for the KPI equation \cite{BPP2006b}. In this framework it was possible to develop the inverse scattering transform for a solution describing one soliton on a generic background~\cite{BPPP2002}, and to study the existence of the (extended) resolvent for (some) multisoliton solutions~\cite{BPPP2009}. However, the general case of $N$-solitons is still open. Following~\cite{BPP2006b}, the first step in building the inverse scattering theory for this case lies in building a Green's function $\Go(x,x',\bk)$ of the heat operator (\ref{heatop}) corresponding to the pure soliton potential  $u(x)$, that is, satisfying
\begin{equation}
\bigl(-\partial_{x_{2}}^{}+\partial_{x_{1}}^{2}-u(x)\bigr)\Go(x,x',\bk)=\delta (x-x'),  \label{green}
\end{equation}
and such that
\begin{equation}
G(x,x',\bk)=e_{}^{i\bk(x_{1}-x_{1}')+\bk^{2}(x_{2}-x_{2}')}\Go(x,x',\bk)  \label{Green}
\end{equation}
is bounded with   respect to the variables $x,x'\in \Rs^{2}$ and $\bk\in\Cs$ and has finite limits at infinity. The construction of such function is the subject of this article.

\section{Heat operator with pure soliton potential and its Jost solutions}

Soliton potentials (see \cite{BPPP2009}--\cite{asympJostKPII} for details) are labeled by the two numbers (topological charges) $N_{a}$ and $N_{b}$, which obey condition
\begin{equation}
N_{a},N_{b}\geq 1.  \label{nanb}
\end{equation}
Let
\begin{equation}
\No=N_{a}+N_{b},  \label{Nnanb}
\end{equation}
so that $\No\geq 2$. We introduce the $\No$ real parameters
\begin{equation}
\kappa_{1}<\kappa_{2}<\ldots <\kappa_{\No},  \label{kappas}
\end{equation}
and the functions
\begin{equation}
K_{n}^{}(x)=\kappa_{n}^{}x_{1}^{}+\kappa_{n}^{2}x_{2}^{},\quad n=1,\ldots ,\No.  \label{Kn}
\end{equation}
Let
\begin{equation}
e^{K(x)}=\diag\{e^{K_{n}(x)}\}_{n=1}^{\No}  \label{eK}
\end{equation}
be a diagonal $\No\times {\No}$ matrix, let $\Do$ be a $\No\times {N_{b}}$ constant matrix and $\Vo$ be an ``incomplete Vandermonde matrix,'' i.e., the $N_{b}\times \No$ matrix
\begin{equation}
\Vo=\left(\begin{array}{lll}
1 & \ldots & 1 \\
\vdots &  & \vdots \\
\kappa_{1}^{N_{b}-1} & \ldots & \kappa_{\No}^{N_{b}-1}
\end{array}\right) .  \label{W}
\end{equation}
Then, the soliton potential is given by
\begin{equation}
u(x)=-2\partial_{x_{1}}^{2}\log \tau (x),  \label{ux}
\end{equation}
where the $\tau$-function can be expressed as
\begin{equation}
\tau (x)=\det\bigl(\Vo e^{K(x)}\Do\bigr).  \label{tau}
\end{equation}
For the Jost and dual Jost solutions we have
\begin{align}
& \Phi (x,\bk)=e^{-i\bk x_{1}-\bk^{2}x_{2}}\chi (x,\bk),  \label{Phi} \\
& \Psi (x,\bk)=e^{i\bk x_{1}+\bk^{2}x_{2}}\xi (x,\bk),  \label{Psi}
\end{align}
where
\begin{align}
& \chi (x,\bk)=\dfrac{\tau_{\Phi}^{}(x,\bk)}{\tau (x)},  \label{symPhi} \\
& \xi (x,\bk)=\dfrac{\tau_{\Psi}^{}(x,\bk)}{\tau (x)},  \label{symPsi}
\end{align}
with (Miwa shift)
\begin{equation}
\tau_{\Phi}(x,\bk)=\det \bigl(\Vo e^{K(x)}(\kappa +i\bk)\Do\bigr),\qquad
\tau_{\Psi}(x,\bk)=\det \left( \Vo\dfrac{e^{K(x)}}{\kappa +i\bk}\Do\right),  \label{tauk:1}
\end{equation}
and
\begin{equation}
\kappa +i\bk=\diag\{\kappa_{n}+i\bk\}_{n=1}^{\No}.
\label{kappadiag}
\end{equation}

In order to study the properties of the potential and the Jost solutions, it is convenient to use an explicit representation for the determinants. By using the Binet--Cauchy formula for the determinant of a product of matrices we get
\begin{align}
& \tau (x)=\dfrac{1}{N_{b}!}\sum_{\{n_{i}\}=1}^{\No}\Do(\{n_{i}\})V(\{n_{i}\})
\prod_{l=1}^{N_{b}}e^{K_{n_{l}}(x)},  \label{tauf1} \\
& \chi (x,\bk)=\dfrac{1}{N_{b}!\tau (x)}\sum_{\{m_{i}\}=1}^{\No}\Do(\{m_{i}\})V(\{m_{i}\})
\prod_{l=1}^{N_{b}}(\kappa_{m_{l}}+i\bk)e_{}^{K_{m_{l}}(x)},  \label{Phik1} \\
& \xi (x,\bk)=\dfrac{1}{N_{b}!\tau (x)}\sum_{\{n_{i}\}=1}^{\No}\Do(\{n_{i}\})V(\{n_{i}\})
\prod_{l=1}^{N_{b}}\dfrac{e_{}^{K_{n_{l}}(x)}}{\kappa_{n_{l}}+i\bk},
\label{Psik1}
\end{align}
where we used notation
\begin{align}
& V(\{n_{i}\})=\det \left(\begin{array}{lll}
1 & \ldots & 1 \\
\vdots &  & \vdots \\
\kappa_{n_{1}}^{N_{b}-1} & \ldots & \kappa_{n_{N_b}}^{N_{b}-1}
\end{array}\right)
\equiv \prod_{1\leq i<j\leq N_{b}}(\kappa_{n_{j}}-\kappa_{n_{i}}),\label{V} \\
& \Do(\{n_{i}\})=\det \left(
\begin{array}{ccc}
\Do_{n_{1},1} & \dots & \Do_{n_{1},N_{b}} \\
\vdots &  & \vdots \\
\Do_{n_{N_{b}},1} & \dots & \Do_{n_{N_{b}},N_{b}}
\end{array}\right)  \label{Do1}
\end{align}
for the maximal minors of matrices $\Vo$ and $\Do$ and where
\begin{equation}
\{m_{i}\}=\{m_{1},\ldots,m_{N_{b}}\},\qquad\{n_{i}\}=\{n_{1},\ldots ,n_{N_{b}}\}  \label{not}
\end{equation}
stand for non ordered sets of $N_b$ indices from the interval $1,\ldots,\No$.

Notice that the only $x$-dependent terms in (\ref{tauf1}), (\ref{Phik1}),  and (\ref{Psik1}) are exponents of sums of linear functions (\ref{Kn}). Correspondingly, the asymptotic behavior of the function $\tau(x)$ and of the potential has a sectorial structure on the $x$-plane. In order to specify these sectors at $x\to\infty$ we introduce the ray directions
\begin{equation}\label{rn}
r_{n}:\qquad\left\{\begin{array}{l}
x_1+(\kappa_{n}+\kappa_{n+N_b})x_2\quad\text{bounded}\\
(\kappa_{n+N_b}-\kappa_{n})x_2\to-\infty.
\end{array}\right.,\quad n=1,\ldots,\No,
\end{equation}
where we assume that the indices are defined mod$\,\No$, so that thanks to (\ref{Nnanb}), say, $n+N_b=n-N_a$ for $n>N_a$. Thus there are $N_a$ rays in the direction $x_2\to-\infty$ and $N_b$ rays in the direction $x_2\to+\infty$. The sector $\sigma_{n}$ is swept out by rotating  anticlockwise the ray $r_n$ up to the ray $r_{n+1}$. These sectors are nonintersecting and cover the whole $x$-plane with the exception of rays. In \cite{asympKPII} we proved that the the leading exponents of $\tau(x)$ when $x\to\infty$ are the exponents $\exp\bigl(\sum_{l=n}^{n+N_b-1}K_{l}(x)\bigr)$, each being the leading one in the corresponding $\sigma_{n}$ sector of the $x$-plane. More exactly, if the coefficients
\begin{equation}
z_{n}=V(\kappa _{n}^{},\ldots ,\kappa_{n+N_{b}-1}^{})\Do(n,\ldots ,n+N_{b}-1),  \label{zn}
\end{equation}
are different from zero for all $n=1,\ldots,\No$ (again with indices defined mod$\,\No$) the function $\tau(x)$ has the following asymptotic behavior along rays and inside sectors:
\begin{align}
&x\stackrel{r_{n}}{\longrightarrow}\infty:&&
\tau (x)=\bigl(z_{n}+z_{n+1}e_{}^{K_{N_{b}+n}(x)-K_{n}(x)}+o(1)\bigr)\exp\Biggl(\sum_{l=n}^{n+N_{b}-1}K_{l}(x)
\Biggr), \label{3:232}\\
&x\stackrel{\sigma_{n}}{\longrightarrow}\infty:&&
\tau(x)=\bigl(z_{n}+o(1)\bigr)\exp\Biggl(\sum_{l=n}^{n+N_{b}-1}K_{l}(x)\Biggr). \label{tauasympt}
\end{align}
Regularity of the potential $u(x)$ on the $x$-plane is equivalent to the absence of zeroes of $\tau(x)$. It is clear that it is enough to impose the condition that the matrix $\Do$  is Totally Non Negative (TNN), i.e., that
\begin{equation}
\Do(n_1,\ldots,n_{N_b})\geq0,\quad\text{for all}\quad1\leq n_{1}<\ldots <n_{N_{b}}\leq\No.\label{D}
\end{equation}
However, sufficient conditions on the matrix $\Do$ for the regularity of the potential are unknown. On the other side, from (\ref{3:232}) and (\ref{tauasympt}) it follows directly that it is sufficient to require that
\begin{equation}
z_n>0\label{zn0}
\end{equation}
for having nonsingular asymptotics of the potential.

We also mention that the functions  $\chi(x,\bk)$ and $\xi(x,\bk)$ have bounded asymptotics on the $x$-planes because the $x$-dependent exponents enter in denominators and numerators of expressions (\ref{Phik1}) and  (\ref{Psik1}) with coefficients proportional to $\Do(\{n_{i}\})$. This means that the leading asymptotic behavior of the denominators of the functions $\chi(x,\bk)$ and $\xi(x,\bk)$ on the $x$-plane is not weaker then the behavior of their numerators. For more details see~\cite{BPPPr2001a}, the review papers \cite{ChK2,K}, and \cite{equivKPII,asympKPII,asympJostKPII}, where the same notations have been used.

We need in the following also the values $\chi (x,i\kappa_{n})$ of $\chi(x,\bk)$ at $\bk=i\kappa_{n}$ and the residues $\xi_{n}(x)$ of $\xi (x,\bk)$ at $\bk=i\kappa_{n}$. From (\ref{Phi}), (\ref{Psi}) and (\ref{Phik1}), (\ref{Psik1}) we have
\begin{align}
&\chi(x,i\kappa_{n})=\dfrac{(-1)^{N_b}}{N_{b}!\tau(x)}\sum_{\{m_{i}\}=1}^{\No}
\Do(\{m_{i}\})V(\{m_{i}\},n)\prod_{l=1}^{N_{b}}e^{K_{m_{l}}(x)},  \label{1} \\
&\xi_{n}(x)=\dfrac{1}{iN_{b}!\tau(x)}\sum_{\{n_{i}\}=1}^{\No}
\Do(\{n_{i}\})\sum_{j=1}^{N_{b}}\delta_{n_{j}n}(-1)^{j-1}V(n_1,\ldots,\widehat{n_j},\ldots,n_{N_b})
\prod_{l=1}^{N_{b}}e^{K_{n_{l}}(x)},  \label{2}
\end{align}
where $\{\{m_{i}\},n\}=\{m_{1},\ldots,m_{N_{b}},n\}$, hat over $n_{j}$ denotes that this index is omitted and where the $\delta_{n_{j}n}$ Kronecker symbol in the r.h.s.\ of the last formula is due to the fact that the residues of the terms in the sum are nonzero only when some $n_{j}=n$.

Taking into account the analyticity properties of $\chi (x,\bk)$ and $\xi (x,\bk)$ in (\ref{Phik1}), (\ref{Psik1}) their product can be written in terms of the values $\chi (x,i\kappa_{n})$ and $\xi_{n}(x)$ as follows
\begin{equation}
\chi (x,\bk)\xi (x',\bk)=1+\sum_{n=1}^{\mathcal{N}}\dfrac{\chi (x,i\kappa_{n})\xi_{n}(x')}{\bk-i\kappa_{n}},\label{chixi}
\end{equation}
which also will be useful in the following. In \cite{asympJostKPII} we demonstrated that the Jost solutions obey the Hirota bilinear identity
\begin{equation}
\sum_{n=1}^{\mathcal{N}}\Phi (x,i\kappa_{n})\Psi_{n}(x')=0,  \label{sumPhiPsi}
\end{equation}
where in analogy to (\ref{2}) $\Psi_{n}(x)$ denotes the residue of $\Psi(x,\bk)$.

\section{Green's function}

We want, now, to prove that
\begin{align}
\Go(x,x',\bk)& =-\dfrac{\sgn(x_{2}-x_{2}')}{2\pi}\int
\!\!ds\,\theta \bigl((s^{2}-\bk_{\Re}^{2})(x_{2}^{}-x_{2}')\bigr)\Phi (x,s+i\bk_{\Im})\Psi (x',s+i\bk_{\Im})+  \notag \\
& +i\theta (x_{2}'-x_{2}^{})\sum_{n=1}^{\No}\theta (\bk_{\Im}-\kappa_{n})\Phi (x,i\kappa_{n})\Psi_{n}(x'), \label{g3}
\end{align}
where $\theta$ denotes the step function of its argument,
satisfies (\ref{green}) and that $G(x,x',\bk)$ defined in (\ref{Green}) is bounded with respect to the variables $x,x'\in \Rs^{2}$ and $\bk\in\Cs$ and has finite limits at space infinity.

Notice that convergence of the integral in (\ref{g3}) follows directly from boundedness of functions $\chi (x,\bk)$ and $\xi (x',\bk)$.
Applying the heat operator (\ref{heatop}) to $\Go(x,x',\bk)$ in (\ref{g3}) we get
\begin{align}
\mathcal{L}(x,\partial_{x})\Go(x,x',\bk)=&\dfrac{\delta(x_{2}-x_{2}')}{2\pi}\int \!\!ds\ \Phi (x,s+i\bk_{\Im})
\Psi(x',s+i\bk_{\Im})+  \notag \\
& +i\delta (x_{2}-x_{2}')\sum_{n=1}^{\No}\theta (\bk_{\Im}-\kappa_{n})\Phi (x,i\kappa_{n})\Psi_{n}(x').  \label{g31}
\end{align}
The integral, after inserting (\ref{chixi}) in it, can be explicitly computed getting
\begin{align}
\dfrac{\delta (x_{2}-x_{2}')}{2\pi}& \int \!\!ds\ \Phi (x,s+i\bk_{\Im})\Psi (x',s+i\bk_{\Im})=\delta (x-x')-  \notag \\
& -i\delta (x_{2}-x_{2}')\sum_{n=1}^{\No}\theta (\bk_{\Im}-\kappa_{n})\Phi (x,i\kappa_{n})\Psi_{n}(x')  \notag \\
& +i\delta (x_{2}-x_{2}')\theta (x_{1}'-x^{}_{1})\sum_{n=1}^{\No}\Phi (x,i\kappa_{n})\Psi_{n}(x').  \label{g32}
\end{align}
Then, thanks to (\ref{sumPhiPsi}), equation (\ref{green}) is proved.

In order to prove boundedness of $G(x,x',\bk)$ we write
\begin{equation}
G(x,x',\bk)=G_{\text{c}}^{}(x,x',\bk)+G_{\text{d}}(x,x',\bk),  \label{g4}
\end{equation}
where the ``continuous'' part $G_{\text{c}}^{}(x,x',\bk)$ of the Green's function equals
\begin{align}
G_{\text{c}}^{}(x,x',\bk)&=-\dfrac{\sgn(x_{2}-x_{2}')}{2\pi}\int\!\!ds\,
\theta\bigl((s^{2}-\bk_{\Re}^{2})(x_{2}^{}-x_{2}')\bigr)
e_{}^{i(\bk_{\Re}-s)[x_{1}^{}-x_{1}'+2\bk_{\Im}(x_{2}^{}-x'_{2})]}\times  \notag \\
&\times e_{}^{(\bk_{\Re}^{2}-s^{2})(x_{2}^{}-x_{2}')}\chi(x,s+i\bk_{\Im})\xi (x',s+i\bk_{\Im}).  \label{g5}
\end{align}
and its ``discrete'' part $G_{\text{d}}(x,x',\bk)$ is given by
\begin{equation}
G_{\text{d}}(x,x',\bk)=i\theta (x_{2}'-x_{2}^{})e^{i\bk(x_{1}-x_{1}')+\bk^{2}(x_{2}-x_{2}')}
\sum_{n=1}^{\No}\theta (\bk_{\Im}-\kappa_{n})\Phi (x,i\kappa_{n})\Psi_{n}(x').\label{g6}
\end{equation}
While the boundedness of $G_{\text{c}}^{}(x,x',\bk)$\ in (\ref{g5}) for $x$, $x'$, and $\bk$ (including limits at infinity) follows directly from boundedness the product $\chi(x,\bk)\xi(x',\bk)$ at infinity, properties of $G_{\text{d}}(x,x',\bk)$ in (\ref{g6}) deserves a more accurate investigation.

Let us consider the sum in the r.h.s.\ of (\ref{g6}). Inserting in it (\ref{1}) and (\ref{2}), thanks to antisymmetry of minors of matrices $\Do$ and $\Vo$ (see (\ref{V}) and (\ref{Do1})), after summing over $n$, we get
\begin{align}
& \sum_{n=1}^{\No}\theta (\bk_{\Im}-\kappa_{n})\Phi (x,i\kappa_{n})\Psi_{n}(x')=  \notag \\
&\qquad=\dfrac{i}{N_{b}!(N_b-1)!\tau(x)\tau(x')}\sum_{\{m_{i}\}=1}^{\No}\sum_{\{n_{i}\}=1}^{\No}
\Do(\{m_{i}\})\Do(\{n_{i}\})\theta (\bk_{\Im}-\kappa_{n_{N_b}^{}})\times  \notag \\
& \qquad \times V(\{m_{i}\},n^{}_{N_b})V(n_{1},\ldots,n_{N_{b}-1}^{})\times\notag \\
& \qquad\times\exp \Biggl(\sum_{l=1}^{N_{b}}K_{m_{l}}(x)+K_{n^{}_{N_b}}(x)+
\sum_{l=1}^{N_{b}-1}K_{n_{l}}(x')\Biggr).\label{4}
\end{align}

We recall, now, that the maximal minors of a matrix satisfy the Pl\"{u}cker relation, i.e., for any subsets $\{m_{i}\}$ and $\{n_{i}\}$ of indices running from $1$ to $\mathcal{N}$ and arbitrary $j\in \{1,\ldots ,N_{b}\}$
\begin{align}
& \Do(\{m_{i}\})\Do(\{n_{i}\})=  \notag \\
& =\sum_{s=1}^{N_{b}}\Do(m_{1},\ldots ,m_{s-1},n_{j},m_{s+1}\ldots,m_{N_{b}})
\Do(n_{1},\ldots ,n_{j-1},m_{s},n_{j+1},\ldots ,n_{N_{b}}).\label{plu}
\end{align}
Inserting this equality with $j=N_b$ in the r.h.s.\ of (\ref{4}) we exchange $m_{s}\leftrightarrow{n_{N_b}}$ for $s=1,\ldots,N_{b}$. Notice that under this transformation the first Vandermonde determinant changes sign, while the second Vandermonde determinant is unchanged, as well as the exponent. Thus
\begin{align}
& \sum_{n=1}^{\No}\theta (\bk_{\Im}-\kappa_{n})\Phi (x,i\kappa_{n})\Psi_{n}(x')=  \notag \\
& \quad =\dfrac{-i}{N_{b}!(N_{b}-1)!\tau (x)\tau (x')}\sum_{s=1}^{N_{b}}
\sum_{\{m_{i}\}=1}^{\No}\sum_{\{n_{i}\}=1}^{\No}\Do(\{m_{i}\})\Do(\{n_{i}\})\times  \notag \\
& \quad\times\theta(\bk_{\Im}-\kappa_{m_{s}})V(\{m_{i}\},n_{N_b})
V(n_{1},\ldots,n_{N_{b}-1})\times  \notag \\
& \quad\times\exp \Biggl(\sum_{l=1}^{N_{b}}K_{m_{l}}(x)+K_{n^{}_{N_b}}(x)+\sum
_{l=1}^{N_{b}-1}K_{n_{l}}(x')\Biggr).\label{5}
\end{align}
Exchanging now $m_s\leftrightarrow{m_{N_b}^{}}$ we get, summing over $s$, $N_b$ equal terms. Finally, we multiply (\ref{4}) by $N_b$, sum up with (\ref{5}), divide this sum by $N_{b}+1$ and insert the result into (\ref{g6}). This gives us
\begin{align}
& G_{\text{d}}(x,x',\bk)=  \notag \\
&\quad=\dfrac{\theta(x_{2}'-x_{2}^{})}{((N_{b}-1)!)^{2}(N_{b}+1)\tau(x)\tau(x')}\sum_{\{m_{i}\}=1}^{\No}
\sum_{\{n_{i}\}=1}^{\No}\Do(\{m_{i}\})\Do(\{n_{i}\})\theta(k^{}_{m_{N_b}n_{N_b}})\times  \notag\\
& \quad \times[\theta (\bk_{\Im}-\kappa^{}_{m_{N_b}})-\theta (\bk_{\Im}-\kappa^{}_{n_{N_b}})]
V(\{m_{i}\},n^{}_{N_b})V(n_{1},\ldots,n^{}_{N_{b}-1})
\times  \notag \\
& \quad \times \exp \Biggl({\sum_{l=1}^{N_{b}}K_{m_{l}}(x)+K_{n^{}_{N_b}}(x)+
\sum_{l=1}^{N_{b}-1}K_{n_{l}}(x')+i\bk(x_{1}^{}-x_{1}')+\bk^{2}(x_{2}^{}-x_{2}'})\Biggr),\label{g7}
\end{align}
where
\begin{equation}
k_{mn}=\bk_{\Re}^{2}-(\bk_{\Im}-\kappa^{}_{m})(\bk_{\Im}-\kappa^{}_{n}),  \label{g7'}
\end{equation}
and, for convenience, we introduced the factor $\theta(k_{m,n})$, which identically equals $1$ for $\bk_{\Im}$ in between $\kappa_{m}$ and $\kappa_{n}$, since $k_{m,n}$ is positive inside this interval.
Let us also denote
\begin{equation}
z_{mn}=x_{1}+(\kappa_{m}+\kappa_{n})x_{2},  \label{zmn}
\end{equation}
and analogously for $z'$ and $x'$, and let us notice that
\begin{align}
\theta (\bk_{\Im}&-\kappa_{m})- \theta(\bk_{\Im}-\kappa_{n})=
\sgn(z_{mn}-z'_{mn})\times  \notag \\
& \times [\theta ((\bk_{\Im}-\kappa_{m})(z_{mn}^{}-z_{mn}'))-\theta ((\bk_{\Im}-\kappa_{n})(z_{mn}^{}-z_{mn}'))].  \label{theta-theta}
\end{align}
and that
\begin{align}
i\bk(x_{1}^{}-x_{1}')+{\bk^{2}(x_{2}^{}-x_{2}'})&+K_{m}(x)=
k_{mn}(x_{2}^{}-x_{2}')-(\bk_{\Im}-\kappa_{m})(z_{mn}-z_{mn}')+  \notag \\
& +i\bk_{\Re}[x_{1}^{}-x_{1}'+2\bk_{\Im}(x_{2}^{}-x_{2}')]+K_{m}(x').  \label{kmn1}
\end{align}

Then, we can decompose $G_{\text{d}}(x,x',\bk)$ in the sum
\begin{equation}
G_{\text{d}}(x,x',\bk)=G_{1}(x,x',\bk)+G_{2}(x,x',\bk),  \label{g8}
\end{equation}
with
\begin{align}
& G_{1}(x,x',\bk)=\notag\\
& \quad =-\dfrac{e^{i\bk_{\Re}[x_{1}^{}-x_{1}'+2\bk_{\Im}(x_{2}^{}-x_{2}')]}}{((N_{b}-1)!)^2(N_{b}+1)}
\sum_{\{m_{i}\}=1}^{\No}\sum_{\{n_{i}\}=1}^{\No}\sgn(z_{m_{N_b}n_{N_b}}-z_{m_{N_b}n_{N_b}}')\times  \notag \\
& \quad \times \theta \bigl(k_{m^{}_{N_b}n^{}_{N_b}}(x_{2}'-x_{2}^{})\bigr)
\theta\bigl((\bk_{\Im}-\kappa_{n^{}_{N_b}})(z_{m_{N_b}n_{N_b}}-z_{m_{N_b}n_{N_b}}')\bigr)\times  \notag \\
&\quad\times e_{}^{k_{m_{N_b}n_{N_b}}(x_{2}^{}-x_{2}')-(\bk_{\Im}-\kappa_{n_{N_b}})(z_{m_{N_b}n_{N_b}}-z_{m_{N_b}n_{N_b}}')}
V(\{m_{i}\},n_{N_b})V(n_{1},\ldots,n_{N_{b}-1})\times  \notag \\
& \quad \times \dfrac{\Do(\{m_{i}\})\exp\displaystyle\Biggl(\sum_{l=1}^{N_{b}}K_{m_{l}}(x)\Biggr)}{\tau (x)}\times \dfrac{\Do(\{n_{i}\})\exp\displaystyle\Biggl(\sum_{l=1}^{N_{b}}K_{n_{l}}(x')\Biggr)}{\tau (x')},
\label{g12}
\end{align}
and
\begin{align}
& G_{2}(x,x',\bk)=  \notag \\
& \quad =\dfrac{e_{}^{i\bk_{\Re}[x_{1}^{}-x_{1}'+2\bk_{\Im}(x_{2}^{}-x_{2}')]}}{((N_{b}-1)!)^2(N_{b}+1)}
\sum_{\{m_{i}\}=1}^{\No}\sum_{\{n_{i}\}=1}^{\No}\sgn(z_{m_{N_b}n_{N_b}}-z_{m_{N_b}n_{N_b}}')\times\notag \\
& \quad \times\theta\bigl(k_{m_{N_b}n_{N_b}}(x_{2}'-x_{2}^{})\bigr)
\theta\bigl((\bk_{\Im}-\kappa_{m_{N_b}})(z_{m_{N_b}n_{N_b}}-z_{m_{N_b}n_{N_b}}')\bigr)\times  \notag \\
& \quad \times e^{k_{m_{N_b}n_{N_b}}(x_{2}^{}-x_{2}')-(\bk_{\Im}-\kappa_{m_{N_b}})(z_{m_{N_b}n_{N_b}}-z_{m_{N_b}n_{N_b}}')}
\times  \notag \\
& \quad \times V(\{m_{i}\},n_{N_b})\dfrac{\Do(\{m_{i}\})
\exp\displaystyle\Biggl(\sum_{l=1}^{N_{b}-1}K_{m_{l}}(x)+K_{n_{N_b}}(x)\Biggr)}{\tau (x)}
\times\notag \\
& \quad \times V(n_{1},\ldots,n_{N_{b}-1}) \dfrac{\Do(\{n_{i}\})\exp\displaystyle\Biggl(\sum_{l=1}^{N_{b}-1}K_{n_{l}}(x')+K_{m_{N_b}}(x')\Biggr)}{\tau(x')}.\label{g11}
\end{align}
The exponents in the third lines of the r.h.s.\ of (\ref{g12}) and (\ref{g11}) are decaying or at least not growing thanks to the $\theta$-functions in the second lines. So we have to check behavior with respect to $x$ and $x'$ of the last lines of these equations. About (\ref{g12}) the situation is trivial. The two exponents in the forth line has the same coefficient (minor of $\Do$) as
in $\tau (x)$ and $\tau (x')$ in the denominator and, therefore, they are bounded when $x$ and $x'$ are growing. Situation with (\ref{g11}) is more involved. Let us consider the term in the forth line. If the minor $\Do(\{m_{i}\})$ in the numerator is different from zero and $\Do(n_{1},\ldots,n_{N_{b}-1})\neq 0$, then the same exponent as in the numerator is present in $\tau(x)$ in the
denominator and the ratio is bounded. But if $\Do(n_{1},\ldots,n_{N_{b}-1})=0$ such exponent is not involved in $\tau (x)$ and in the direction where it is the leading one (if such direction exists) the ratio is growing at large space. The same is valid for the term in the fifth line of (\ref{g11}). Thus, in the case of a Totally Positive matrix $\Do$ the boundedness of the Green's function $G(x,x',\bk)$ (\ref{Green}) is proved. If instead of TP we impose on the matrix $\Do$ conditions (\ref{zn0}), then all leading exponents are involved in $\tau(x)$, as we mentioned in discussion of (\ref{3:232}) and (\ref{tauasympt}). Thus again, both ratios in (\ref{g11}) have finite asymptotics at space infinity, but can be singular in the finite domain. To avoid this singularities it is enough to impose additionally that the matrix $\Do$ is TNN. But the case of a generic TNN matrix $\Do$ requires an additional study.

Boundedness of the Green's function $G(x,x',\bk)$ with respect to the variable $\bk$ also follows from (\ref{g5}), (\ref{g12}) and (\ref{g11}). In all points $\bk=i\kappa_{1},\ldots,i\kappa_{\No}$ this function is discontinuous. As it was shown in \cite{BPPP2002}, in the case $N_a=N_b=1$, in order to give the detailed behavior of the Green's function at the points of discontinuity and in order to determine possible additional Green's functions useful in defining spectral data and studying their properties one needs a more generic object: the extended resolvent of the heat operator. Here, in the following section, we introduce the extended resolvent, leaving for a forthcoming article the study of its properties (much more involved than in the case of the nonstationary Schr\"odingher operator) and its use in building the inverse scattering theory.

\section{Extended resolvent}

Let us recall that the resolvent $M(q)$ of the extended heat operator $L(q)$ which has kernel
\begin{equation}
L(x,x';q)=\mathcal{L}(x,\partial_{x}+q)\delta (x-x'),\qquad q=(q_{1},q_{2})\in \mathbb{R}^{2},  \label{Lext}
\end{equation}
is defined as the operator with kernel $M(x,x';q)$, which is the left and right inverse of $L(q)$.

In \cite{asympJostKPII} we showed that for a pure soliton potential the operator $L(q)$ has left annihilators if $N_{a}<N_{b}$ for $q$ belonging to the union of some polygons included in the polygon inscribed in the parabola $q_{2}=q_{1}^{2}$ of the $q$-plane, with vertices at the points $(\kappa_{n},\kappa_{n}^{2})$, $n=1,\ldots,\No$ and, analogously, we showed
that, for $N_{a}>N_{b}$, the operator $L(q)$ has right annihilators in an analogous region of the $q$-plane. Therefore, we need to find an operator $M(q)$ with bounded kernel $M(x,x';q)$ defined for any $q\in \mathbb{R}^{2}$, which, for $N_{a}<N_{b}$, is right inverse of $L(q)$ for any $q$, but left inverse only for $q$ not belonging to the union of polygons quoted
above, and which, for $N_{a}>N_{b}$, is left inverse of $L(q)$ for any $q$, but right inverse only for $q$ not belonging to an analogous union of polygons. Here we give only the result and we leave the lengthy and involved deduction to a following paper.

The searched operator $M(q)$ has kernel
\begin{equation}
M(x,x';q)=M_{\text{c}}(x,x';q)+M_{\text{d}}(x,x';q),  \label{M}
\end{equation}
where
\begin{align}
M_{\text{c}}(x,x';q)& =-\sgn(x_{2}-x_{2}')\dfrac{e^{-q(x-x')}}{2\pi}\int \!\!dp_{1}\theta\bigl((q_{2}+p_{1}^{2}-q_{1}^{2})(x_{2}-x_{2}')\bigr)\times   \notag \\
& \times \Phi (x;p_{1}+iq_{1})\Psi (x';p_{1}+iq_{1}),  \label{Mc}
\end{align}
and
\begin{align}
& M_{\text{d}}(x,x';q)=\dfrac{\sgn(x'_{2}-x_{2})e^{-q(x-x')}}{((N_{b}-1)!)^2(N_{b}+1)!\tau (x)
\tau (x')}\sum_{\{m_{i}\}=1}^{\No}\sum_{\{n_{i}\}=1}^{\No}\Do(\{m_{i}\})\Do(\{n_{i}\})\times   \notag \\
& \quad \times [\theta (q_{1}-\kappa_{m_{N_b}})-\theta (q_{1}-\kappa_{n_{N_b}})]
\theta \bigl(q_{m_{N_b}n_{N_b}}(x_{2}-x_{2}')\bigr)\times\notag \\
& \quad \times V(\{m_{i}\},n_{N_b})V(n_{1},\ldots ,n_{N_{b}-1})\times   \notag \\
& \quad \times \exp \Biggl({\sum_{l=1}^{N_{b}}K_{m_{l}}(x)+K_{n_{N_b}}(x)+
\sum_{l=1}^{N_{b}-1}K_{n_{l}}(x')\Biggr)},  \label{Md}
\end{align}
with
\begin{equation}
q_{mn}=q_{2}-(\kappa_{m}+\kappa_{n})q_{1}+\kappa_{m}\kappa_{n}. \label{2A42}
\end{equation}
One can check directly that the Green's function $\Go(x,x',\bk)$ can be obtained via the reduction
\begin{equation}
G(x,x',\bk)=e_{}^{i\bk_{\Re}[x_{1}^{}-x_{1}'+2\bk_{\Im}(x_{2}^{}-x_{2}')]}
M(x,x';\bk_{\Im}^{},\bk_{\Im}^{2}-\bk_{\Re}^{2}).  \label{g2'}
\end{equation}
Notice that the points $(q_{1}=\bk_{\Im}^{},q_{2}=\bk_{\Im}^{2}-\bk_{\Re}^{2})$ lies outside the parabola $q_{2}=q_{1}^{2}$ in accordance with the fact that in this region the operator $L(q)$ has right and left inverse. On
the parabola the only points, where $L(q)$, for $N_{a}\neq N_{b}$, has no inverse are the points $\bk=i\kappa_{n}$, $n=1,\dots,\No$ and, therefore, as expected, the Green's function $G(x,x',\bk)$, being a reduction of the resolvent, is discontinuous just at these points.

\section{Acknowledgments}
This work is supported in part by the grant RFBR \# 11-01-00440, Scientific Schools 4612.2012.1, by the Program of RAS ``Mathematical Methods of the Nonlinear Dynamics,'' by INFN, by MIUR (grant PRIN 2008 ``Geometrical methods in the theory of nonlinear integrable systems'', and by Consortium E.I.N.S.T.E.IN.

\end{document}